\documentclass[12pt]{iopart}
\usepackage{iopams}

\usepackage{graphicx}% Include figure files
\begin{document}

\title{Magnetoquantum Oscillations in the Specific Heat of a Topological Kondo Insulator}

\author{P. G.  LaBarre$^1$, A. Rydh$^2$,J. Palmer-Fortune$^3$,J. A. Frothingham$^3$,S. T. Hannahs$^4$, A. P. Ramirez $^1$, N. Fortune $^3$ }
%\email{pglabarre@gmail.com}

%\author{A. Rydh$^2$}

%\author{J. Palmer-Fortune$^3$}

%\author{J. A. Frothingham$^3$}
% \address{Department of Physics, Smith College, Northampton MA 01063}

%\author{A. Bangura} 
%\affiliation{National High Magnetic Field Laboratory, Florida State % University, Tallahassee, FL 32310-3706, USA}

%\author{S. T. Hannahs$^4$} 
%\address{National High Magnetic Field Laboratory, Florida State University, Tallahassee, FL 32310-3706, USA}

%\author{L. Pressley}
%\affiliation{The Johns Hopkins University, Baltimore, MD, USA}

%\author{T. McQueen}
%\affiliation{The Johns Hopkins University, Baltimore, MD, USA}

%\author{A. P. Ramirez $^1$ }
%\email{apr@ucsc.edu}
% \address{Department of Physics, University of California at Santa Cruz, Santa Cruz, CA, USA}

% \author{N. Fortune $^3$}
%\email{nfortune@smith.edu}
%\address{Department of Physics, Smith College, Northampton MA 01063}

\address{$^1$ Department of Physics, University of California at Santa Cruz, Santa Cruz, CA, USA}
\address{$^2$ Department of Physics, Stockholm University, Stockholm, Sweden}
\address{$^3$ Department of Physics, Smith College, Northampton MA 01063}
\address{$^4$ National High Magnetic Field Laboratory, Florida State University, Tallahassee, FL 32310-3706, USA}

\date{\today}

\begin{abstract}
Surprisingly, magnetoquantum oscillations (MQO) characteristic of a metal with a Fermi surface  have been observed in measurements of the topological Kondo insulator SmB\textsubscript{6}. As these  MQO have only been observed in measurements of magnetic torque (dHvA) and not in measurements of magnetoresistance (SdH), a debate has arisen as to whether the MQO are an extrinsic effect arising from rare-earth impurities, defects, and/or aluminum inclusions or an intrinsic effect revealing the existence of charge-neutral excitations. We report here the first observation of  magnetoquantum oscillations in the low-temperature specific heat of SmB\textsubscript{6}.  The observed frequencies and their angular dependence for these flux-grown samples are consistent with previous results based on magnetic torque for SmB\textsubscript{6} but the inferred effective masses are significantly larger than previously reported. Such oscillations can only be observed if the MQO are of bulk thermodynamic origin; the measured magnetic-field dependent oscillation amplitude and effective mass allow us to rule out suggestions of an extrinsic, aluminium inclusion-based origin for the MQO.    \end{abstract}

\maketitle
Despite five decades of study, the Kondo insulator SmB\textsubscript{6} continues to yield new physics, the most recent discovery being the observation of magneto-quantum oscillations (MQOs) characteristic of metals. The MQO are periodic in inverse magnetic field (1/H) and are field-angle dependent. Curiously, these oscillations are observed in measurements of the magnetic torque \cite{li_two-dimensional_2014,tan_unconventional_2015} but not in charge transport \cite{Wolgast_2017}. 
% This dichotomy has led to suggestions of charge neutral Fermi-liquid excitations \cite{Chowdhury_2018,hartstein_intrinsic_2020}. 
As a result, the physical origin of these oscillations continues to be debated.  Arguments have been put forward in favor of  an extrinsic origin dependent  on impurities, defects, and/or sample growth methods  
\cite{ thomas_quantum_2019, fuhrman_magnetic_2020}, a 2D surface state  \cite{li_two-dimensional_2014, li_emergent_2020} , and a 3D non-conducting metallic state  \cite{tan_unconventional_2015, hartstein_fermi_2018,  hartstein_intrinsic_2020} formed from  a charge-neutral Fermi-liquid   \cite{ Chowdhury_2018}  charge-neutral excitons in a Kondo insulator \cite{Knolle_2015, Knolle_2017}, or Majoranas in a mixed valence insulator \cite{varma_majorana_2020}. 

With so many points still in contention, complementary thermodynamic probes of the quantum oscillations merit consideration. In particular, if the magnetoquantum oscillations do reflect an intrinsic and also bulk 3D thermodynamic property of the material (independent of growth conditions) then these oscillations must also occur in the specific heat of both flux-grown and float zone grown samples  \cite{varma_majorana_2020,  shoenberg_magnetic_1984, sullivan_steady-state_1968}. Indeed, MQO have been previously observed in specific heat measurements of very low carrier density semimetals  \cite{shoenberg_magnetic_1984, sullivan_steady-state_1968} and molecular conductors \cite{fortune_specific-heat_1990, bondarenko_first_2001}.   

For a normal metal, Lifshitz-Kosevich (L-K) theory predicts the magnitude of the quantum oscillations in the heat capacity for ordinary metals to be on the order of 0.1\% of the ordinary electronic specific heat \cite{shoenberg_magnetic_1984}.  Our working assumption is that if the observed oscillatory behavior in SmB\textsubscript{6} arise from regions of the sample that can support large mean free paths of Fermi liquid-like excitations \cite{varma_majorana_2020} then they will still be governed by L-K theory regarding oscillation amplitudes and frequencies,  even if the material itself is an insulator and  the excitations are charge neutral. Despite the resulting expected low signal/noise ratio for these oscillations in such a case, their frequencies are still in principle resolvable using Fourier analysis.  

To investigate this possibility, we have measured the low temperature specific heat of both LaB\textsubscript{6} and SmB\textsubscript{6} as a function of applied magnetic field. For SmB$6$, three separate  flux-grown crystals of 0.43 $\mu$g, 1.511 $\mu$g, and 0.126 $\mu$g were compared, each grown in a separate batch at Los Alamos National Laboratory \cite{PhysRevB.99.045138}.  Our measurements of $C(T,H)$ were carried out using custom-built rotatable micro- and nano-calorimeters \cite{tagliati_differential_2012, fortune_top-loading_2014} at $T < 1$K for magnetic fields up to 32 T and also as a function of temperature between 0.1 K and 100 K in magnetic fields up to 12 T. The heat capacities of the bare calorimeters were measured in separate runs and subtracted from the data; corrections were also made for the magnetoresistance of the thermometers \cite{fortune_high_2000}.

\begin{figure}[htb!]
\begin{center}
\includegraphics[width=0.75\columnwidth]{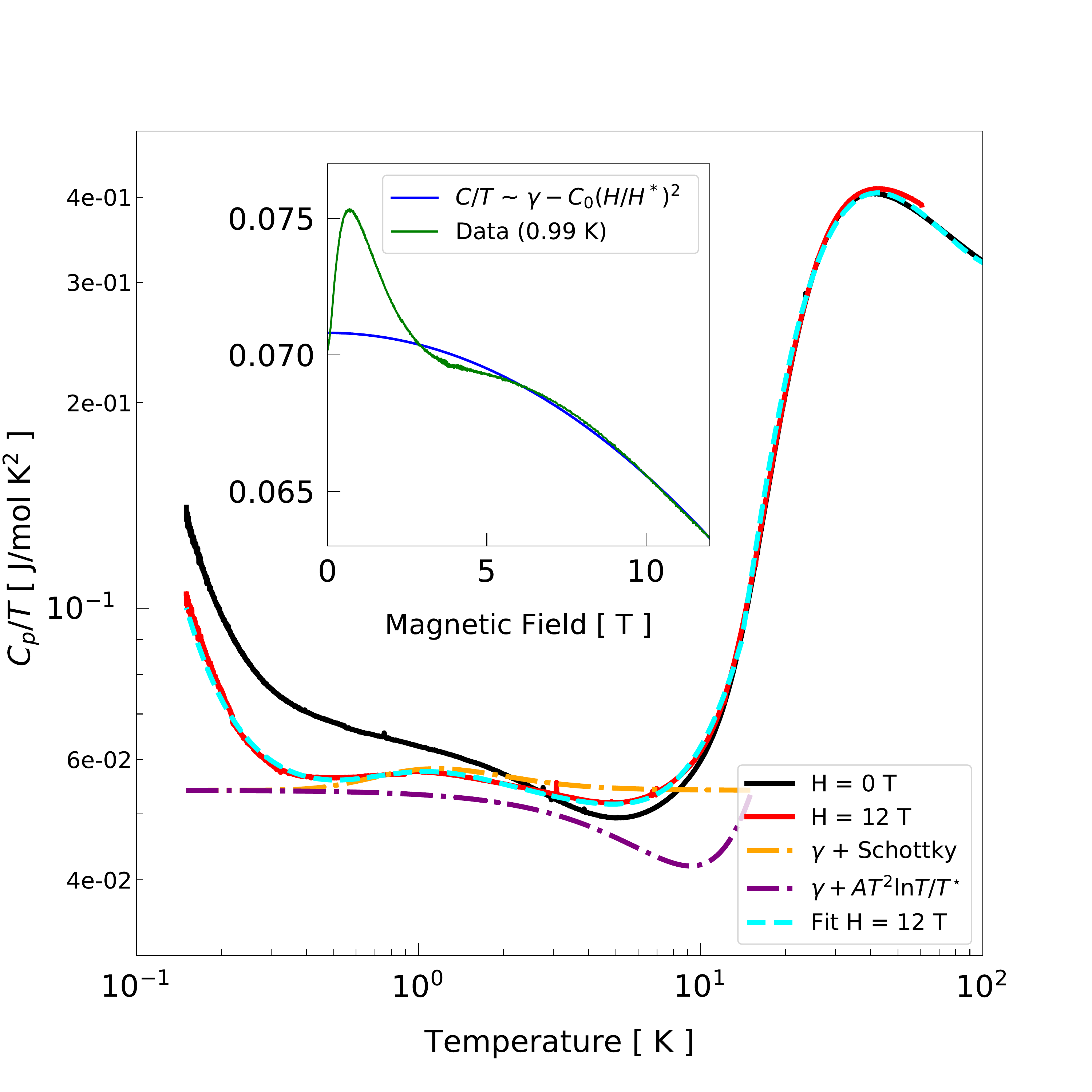}
\caption{\label{fig:zerofield}Temperature dependence of the specific heat at 0 T and 12 T for a 0.430 $\mu$g flux-grown SmB$_6$ sample, along with a representative fit to the temperature dependence at 12 T.   Inset: field dependence of $C/T$ at $T = 0.99$ K. There is a Schottky peak near 1 T arising from magnetic impurities. The $T\ln{T/T^{*}}$ term and the low field $\gamma(0)(1 - \alpha(H/H^{*})^2$ field dependence are consistent with the presence of spin fluctuations with $T^{*}$ = 15 K ($H^{*}$ = 23 T). $\gamma(H)$ is a constant above 28 T (not shown).}
\end{center}
\end{figure}

% Al inclusions also need to be considered as a source of oscillations\cite{thomas_quantum_2019} .
% In addition, flux-grown samples are known to possess inclusions of crystalline Al, the presence of which must be taken into account. 
% Given these caveats, the present pursuit of $C(T)$ in high fields is based on an assumption that any observed oscillatory behavior will be caused by regions of the sample that can support large mean free paths of Fermi liquid excitations. 

In Fig.~\ref{fig:zerofield} we compare zero field and 12 T data for H $\parallel$ a axis, along with a representative fit to the 12 T data for a 0.430 $\mu$g sample. We note in the literature a large variation in the reported low temperature zero-field electronic specific heat $\gamma$ values  for different samples, so our primary interest here is in the magnetic field dependence of $\gamma(H)$ rather than the absolute value.  Such a large sample-to-sample variation is unusual for most materials, but is characteristic of  SmB$_6$  and may be due to the variation in number and type of rare-earth impurities or in the density of mid-gap states \cite{Rosa2021BulkAS}.

Our results at both 0 T and 12 T are well described at low \emph{T} by the following model:
\begin{eqnarray}
\label{eq:model_fit}
C  & =C_{el} + C_{KI} + \beta_{D}T^3 +  DT^{-2} \\
C_{el} &  = \gamma_0 T\ \left[\left(m^{\star }/{m}\right)+A T^2 \ln{\left( {T}/{T^{\star }} \right)} \right] 
\end{eqnarray}
Here $\gamma_0$ in $C_{el}$ is the ``bare'' electronic coefficient of the specific heat expected from band structure, \( m^{\ast}/{m}=\gamma\left(H\right)/{\gamma_0}\) is the many-body effective mass enhancement above the band mass $m$, $A$ is a coupling constant dependent on the strength of the exchange interaction between Fermi-liquid quasiparticles and mass-enhancing excitations, $D/T^2$ is an empirically determined fitting term for the lowest temperature behavior,  
%where $D/T^2$ will be discussed below, 
and \emph{\(T^{\ast}\)} is the characteristic temperature for the excitations \cite{ikeda_quenching_1991}. 

The three non-electronic terms in Eq.~\ref{eq:model_fit} include, first,  a highly sample-dependent Schottky-like term $C$\textsubscript{KI} arising from the temperature dependent screening of magnetic impurities in a Kondo insulator \cite{fuhrman_magnetic_2020}. Numerically, the low and high temperature limits of this model closely match the standard Schottky expression for a two-level system with an energy gap~\(\Delta\) and ground state/excited state degeneracy ratio g\textsubscript{0}/g\textsubscript{1} = 2 \cite{gopal_esr_specific_1966}. We have therefore used the Schottky expression as a proxy for this model (which lacks a numerical prediction for intermediate temperatures).   Second, we include a term \(\beta_{D}T^3\) to represent the low temperature limit of the lattice specific heat in the Debye approximation. Third, we add an empirically fit $DT^{-2}$ term to represent an anomalous  upturn in $C$ with decreasing $T$   \cite{FLACHBART_2002,Gabni_2001,flachbart_specific_2006} analogous to but steeper than previously  seen in heavy fermion systems \cite{FLACHBART_2002,Gabni_2001,flachbart_specific_2006, stewart_non-fermi-liquid_2001}. Nuclear Schottky contributions observed at still lower temperatures in applied magnetic fields  \cite{flachbart_specific_2006} have the same $T^{-2}$ dependence but are considered to be too small to be observed here in our data \cite{hartstein_fermi_2018}.

Turning now to the electronic contributions to the specific heat, we note the growing evidence for intrinsic low temperature magnetism in SmB\textsubscript{6} \cite{gheidi_intrinsic_2019}.  Thus, it is reasonable to expect an additional $T^3 \ln{(T/T^{\star})}$ contribution due to spin fluctuations, as previously observed in other Kondo systems~\cite{ikeda_quenching_1991}, heavy fermions \cite{stewart_possibility_1984} and other electron mass-enhanced metals  \cite{brinkman_spin-fluctuation_1968}.    In SmB\textsubscript{6}, the $T^3\ln(T)$ term has been used to model the dependence of the low temperature specific heat of SmB\textsubscript{6} on carbon doping \cite{phelan_correlation_2014} and (La, Yb) rare earth substitution \cite{orendac_isosbestic_2017}.  Specific heat measurements in a field can therefore provide a critical test: if spin fluctuations are the source of the zero field $T^3 \ln{(T/T^{\star})}$ contribution  and mass enhancement $m^{\star}/m$, that enhancement should be  significantly reduced for fields greater than or on the order of \(H^{\ast}=\frac{k_BT^{\ast}}{\mu_B}\) where $k_BT^*$ is a characteristic energy for spin fluctuations. This reduction results in a decrease in the quasi-particle enhanced effective mass ratio $m^{\star}/m$ and thus $\gamma(H)$,  
%More specifically, 
which should be proportional to $(H/H^{\star})^2$ at low fields \cite{hertel_effect_1980, beal-monod_field_1983}. For $T^{\star}$ = 15 K,  $H^* = 23$ T.
%For the data shown in Fig.~\ref{fig:zerofield}, we find $ T^*$ = 15 K and thus $H^*$ = 23 T. 

Our observations at both low and high temperatures are consistent with the previously observed behavior discussed above.  Consistent with this expectation, we find that  $C/T$ for all measured temperatures ($T\leq 1K$) begins to significantly decrease above 18 T, leveling off above 22 T; the initial field dependence is proportional to $(H/H^*)^2$ (Fig.~\ref{fig:zerofield} inset). A low field Schottky peak around 1-2 T arises from the magnetic impurities present in the sample, as expected; we attribute a second peak seen in higher fields around 12 - 15 T to the experimentally observed suppression of the gap between the in-gap states and the conduction band \cite{caldwell_high-field_2007} by a magnetic field on the order of 14 T.  

%This magnetic-field-dependent gap is on the
%order of 30 K at zero field but is reported to close to zero at a field
%of approximately 14 T ~\hyperref[csl:31]{(\textit{31})}.~~

% \begin{figure}[htb!]
%\begin{center}
%\includegraphics[width=0.70\columnwidth]{figures/Fig2}
%\caption{{Magnetic field dependence of specific heat at 0.99 K.~ A Schottky peak
%in C/T arising from magnetic impurities in the sample is observed at 1
%tesla. The field dependence at low field is proportional to
%1/H\textsuperscript{2}, as expected for a field-dependent
%spin-fluctuation enhanced electronic specific heat                     \left(\frac{m^{\star}}{m}\right)\left(\frac{m}{m_e}\right)
%coefficient~\(\gamma(H)\). The inset shows measurements up to 45 T
%at 0.3 K. At this temperature, an additional Schottky peak appears
%around 15 T; we attribute this to the magnetic-field-induced closing of
%the gap between the ``midgap states'' and the conduction band at this
%same field value \protect\hyperref[csl:31]{(\textit{31})}.
%\end{center}
% \end{figure}
Finally, we consider the oscillatory component of the specific heat. Normally such MQOs arise from the motion of charge carriers and here we are asking, irrespective of the nature of coupling to a gauge field, is there evidence for MQOs in the specific heat and thus the density of states? We show here that the answer is yes. 

 Magnetoquantum oscillations in thermodynamic quantities such as magnetization and specific heat that are periodic in inverse magnetic field arise from oscillations in the thermodynamic potential $\tilde{\Omega} = {\tilde{\Omega}}_0  f_T(z)$, where  $\tilde{\Omega_0}$, is the zero temperature potential,  $f_T(z) = z/\sinh{z}$ is a thermal smearing factor, and  $z \equiv   \pi^2 p \left(m^{\star}/m  \right) \left(m /m_e \right) \left(k_B T/  \mu_B H\right)$ is a dimensionless quantity proportional to effective mass, temperature, and inverse magnetic field. Here \(m^{\star}\) denotes the quasi-particle interaction enhanced mass, \(m\) is the band mass, \(m_e\) is the bare electron mass, and $p$ is an integer denoting the harmonic.

In the standard Lifshitz-Kosevich [L-K]  model of magnetoquantum oscillations for a 3D Fermi surface \cite{shoenberg_magnetic_1984} with extremal area $A$, the potential ${\tilde{\Omega}}_0$ and hence magnetization and specific heat are periodic in inverse field 1/H with a frequency of oscillation $F$ in tesla given by $F = \left(\frac{ \hbar}{2 \pi e}A \right)$. The field and temperature dependent amplitude of the oscillatory specific heat corresponding to the pth harmonic of oscillation frequency $F$ can be written as 
\begin{equation}
\label{eq:c_osc}
\tilde{C}= 2 k_B \left(\frac{e H}{hc}\right)^{3/2}\frac{1}{|A^{\prime\prime}|^{1/2}} zf^{\prime\prime}(z) f_D  p^{-3/2}  \cos\left[2\pi p\left(\frac{F}{H}+\phi  \right)\pm \frac{\pi}{4}\right]
\end{equation}
where $A^{\prime\prime}= | \partial^2 A(E_F)/{\partial k}^2 |$ is a measure of the Fermi surface curvature,   $\phi$ is a phase constant which may have any value between 0 and 1/2, and $f_D$ is the Dingle factor  $f_D = e^{-p \pi^2  \left(k_B T_D^{\star}/ \mu_B  H\right)}$ \cite{shoenberg_magnetic_1984, sullivan_steady-state_1968}.  Here we have adopted the modern   practice \cite{ shoenberg_magnetic_1984} of expressing $f_D$ in terms of an effective mass  $T_D^{\star} = (m^{\star}/ m_e) T_D$ instead of the originally defined band mass temperature $T_D$  \cite{dingle_magnetic_1952-1, dingle_magnetic_1952-2}. Importantly for the interpretation of the data presented here,  $f_T^{\prime\prime}(z) = 0$ at \(z\approx 1.61\), meaning there will be a node in the amplitude of the oscillatory specific heat as a function of magnetic field. Further, the field at which this node is observed is independent of oscillation frequency. As the value of $z$ depends only on the temperature, magnetic field, and effective mass,  the value of the effective mass can be directly determined from the node's location in field \cite{bondarenko_first_2001}.

\begin{figure}[htb!]
\begin{center}
\includegraphics[width=0.75\columnwidth]{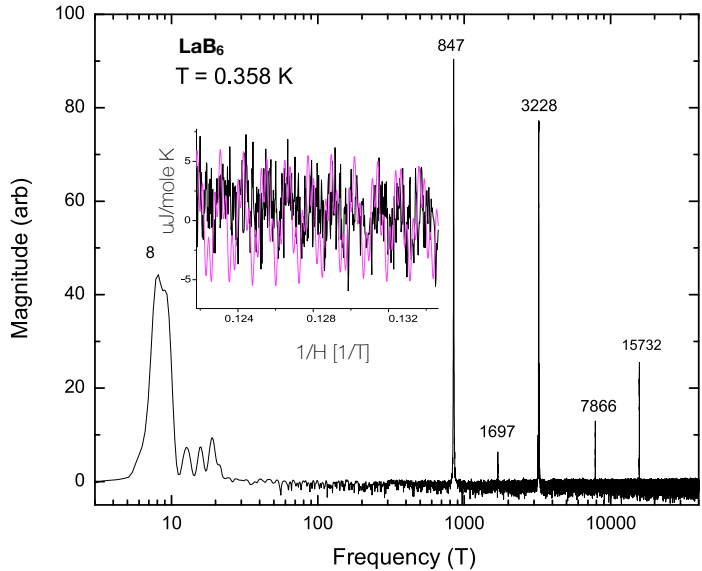}
\caption{{\label{fig:LaB6graph} Fourier power spectrum indicating observed MQO oscillation frequencies for LaB6 in the [001] direction at 0.358 K between 8 and 12 T.  Inset: Inverse field dependence of oscillatory specific heat from which Fourier power spectrum was generated (in black), along with an L-K model fit to the data using the identified frequencies (in purple).
}
}
\end{center}
\end{figure}

The size of the magneto-quantum oscillations in $C(T)$ predicted by this model for LaB$_6$ and SmB$_6$ are much smaller fraction of the total signal than the zero field electronic component of $C(T)$. In the small signal limit, a Fourier transform is needed to pull the signal out of the noise but  to avoid introducing systematic errors in oscillation peak frequency identifications when performing the Fourier transform, it is necessary to use a more general non-uniform discrete Fourier transform (NDFT) technique known as the Lomb-Scargel (LS) method \cite{vanderplas_understanding_2018}. This method, widely used in astronomy, generates the uniform-in-$1/H$ data sets needed for accurate MQO frequency determination in a manner that avoids aliasing errors that would otherwise be introduced by more common interpolation techniques followed by a standard FFT. As with any Fourier transform method, it is also necessary to subtract off a smoothly varying uniform $C(H)$ background. To avoid the introduction of artificial low-frequency peaks, we fit the data to and then subtracted a non-oscillatory sigmoid function (rather than a simple polynomial expansion) before carrying out the frequency analysis.  
% A discussion of our use of the Lomb-Scargle method and the applied peak selection criteria is presented in the supplemental material.

\begin{figure}[ht!]
\begin{center}
\includegraphics[width=0.75\columnwidth]{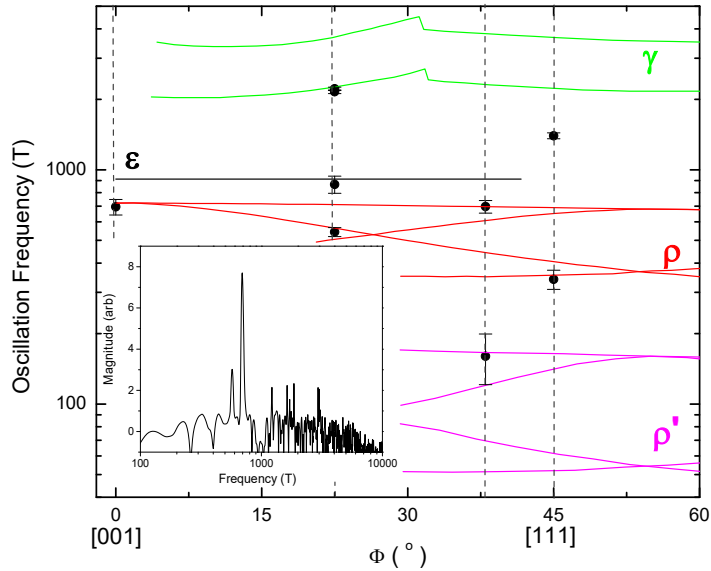}
\caption{{\label{fig:SmB6_angle_dependence} Angular dependence of the oscillation frequencies identified in the Fourier power spectrum for a 1.511 mg flux grown sample of SmB\textsubscript{6} (solid points) in fields up to 32 T. For clarity, we have included only the most prominent peak(s) observed in the Fourier power spectrum at each orientation.Solid lines correspond to predicted angle-dependent oscillation frequencies for SmB\textsubscript{6} \cite{hartstein_fermi_2018};  The band labels $\rho$, $\rho^{\prime}$, $\epsilon$, and $\gamma$ are as presented there.   Inset: a representative Fourier power spectrum for SmB$_6$ with a peak oscillation frequency of 695 T, in this case for a field oriented along [001]. Despite the inherently small size of the oscillatory specific heat for this material, the oscillation frequencies are readily identified, and are in good agreement with magnetic torque measurements and band structure predictions. 
}
}
\end{center}
\end{figure}

As a test of our method, we first measured the oscillatory specific heat of a 0.085 $\mu$g sample of LaB\textsubscript{6}.
%Fig.{\ref{774497}} shows the output of the Lomb-Scargle
%frequency analysis for a field sweep at 0.358 K; the inset shows the
%corresponding fit of the L-K model to the data. 
Applying the Lomb-Scargle frequency analysis described above for a field sweep between 8 and 12 T at 0.358 K, we resolved sharp frequency peaks shown in Fig.~\ref{fig:LaB6graph}:  F = 847($\pm 8$), 1697($\pm 18$), 3228($\pm 15$), 7866($\pm 16$), and 15732($\pm 21$) T, in excellent agreement with previously reported values of 845, 1690, 3220, 7800, and 15600 T based on dHvA measurements \cite{arko_1976, ishizawa_haas-van_1977},  plus a broad peak at approximately 8 T in fair agreement with oscillations reported at 4 T in magnetoresistance \cite{arko_1976} and at 5 T in sound velocity \cite{suzuki_acoustic_1985, thalmeier_effect_1987}. The small amplitude of the oscillations means the signal to noise ratio is low as a function of inverse magnetic field but the oscillation frequencies are readily and accurately identified using the Lomb-Scargle frequency analysis method.

For SmB\textsubscript{6}, we applied the same analysis method as used for LaB\textsubscript{6}. The identified frequencies and their orientation dependence are in good agreement with previous results \cite{hartstein_fermi_2018}. In  Fig. \ref{fig:SmB6_angle_dependence} compare the predicted oscillation frequencies \cite{tan_unconventional_2015}  (shown as green, black, red and pink lines) with frequencies determined from measurements of the oscillatory specific heat made between between 18 and 31 T. For clarity, only  the most prominent peaks in each Fourier power spectrum are shown. As can be seen by the Fourier power spectrum for a magnetic field oriented along [001] shown in the inset to Fig.~\ref{fig:SmB6_angle_dependence},  we are unable to clearly resolve the lowest expected oscillation frequencies in most cases but find an approximately  90\% fidelity agreement at higher frequencies with the values reported for magnetic-torque measurements on float zone grown samples \cite{hartstein_fermi_2018}.

%The agreement in oscillation frequencies determined by different measurement methods on samples also grown by different methods points to a common intrinsic physical origin for the magnetoquantum oscillations, despite their absence in magnetotransport measurements. 

\begin{figure}[htb!]
\includegraphics[width=0.75\columnwidth]{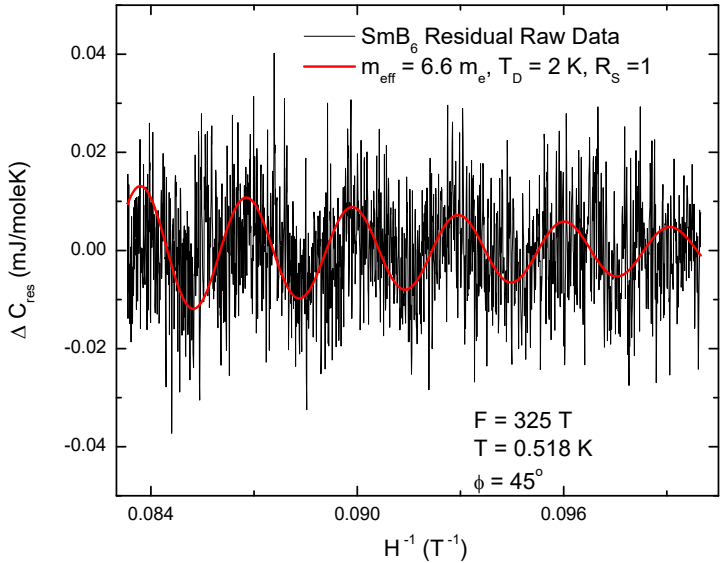}
\caption{\label{fig:membranenanocalorimeter}  Oscillatory specific heat vs $H^{-1}$ at T = 0.518 K for $\Phi = 45^o$ for a field sweep between 10 and 12 T. Data is shown in black. The red curve is the L-K prediction for the most prominent frequency of oscillation in the data (325 T) identified in the Fourier power spectrum at this orientation, with m\textsubscript{eff} = 6.6 m\textsubscript{e} and T\textsubscript{D} = 2 K}.
\end{figure}

For confirmation of the observed  MQOs in $C(H)$ of SmB\textsubscript{6}, a second set of measurements were made on a 0.126 $\mu$g Al-flux grown sample using a high-resolution membrane nanocalorimeter \cite{tagliati_differential_2012}, as shown in Fig.~\ref{fig:membranenanocalorimeter} for a field sweep at 0.52 K for $ H \parallel [111]$ up to 12 T. In this field range, the data is best fit assuming $m^* = 6.6 m_e$. 
 
\begin{figure}[htb!]
\begin{center}
\includegraphics[width=0.75\columnwidth]{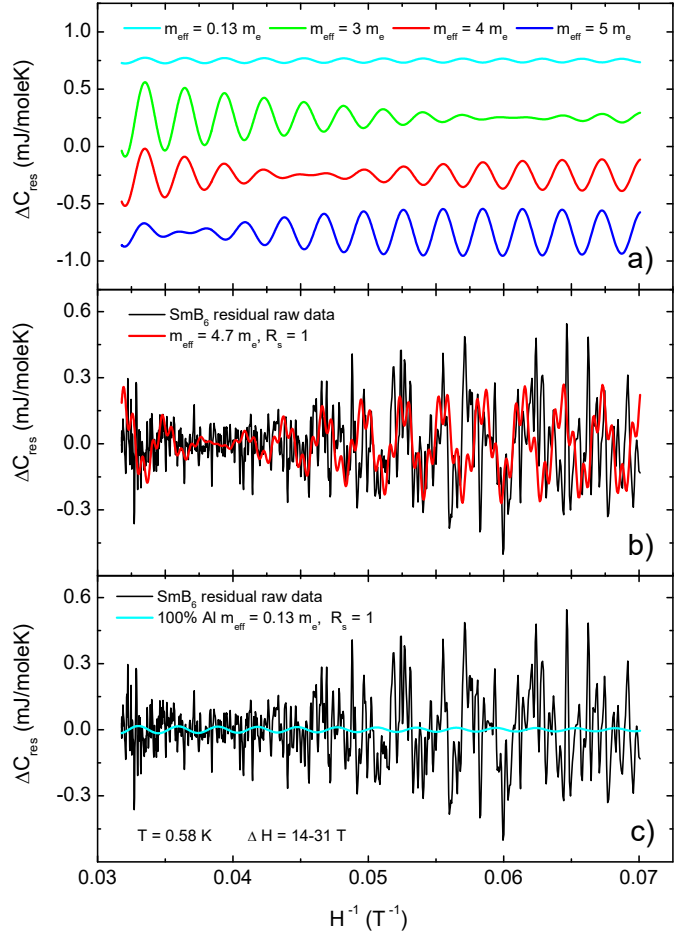}
\caption{{{\label{fig:LKfits}}(a) Illustration of the dependence of the oscillatory specific heat on effective mass, for a sample temperature T = 0.58K, a Dingle temperature $T_D = 1$K and an oscillation frequency F = 341 T, for effective mass enhancement ratios 0.13, 3, 4, and 5 respectively.  Traces  offset vertically for clarity. (b) Residual specific heat vs $H^{-1}$ at T = 0.58K and $\Phi =45^o$. The red curve is  a  comparison of the data to a fit of the L-K model using an effective mass of $4.7 m_e$ determined from the node in the oscillatory specific heat and the two frequencies of oscillation (341 T and 1399 T) identified from the Fourier power spectrum at this orientation.   (c) Comparison of the data with the L-K model prediction for a 100\% Al sample with effective mass of $0.13m_e$ and $T_D = 1$ K. For all fits, we assume a spin splitting factor $R_S = 1$. }}
\end{center}
\end{figure}

To determine the effective mass independent of fitting parameters, we can make use of our observation at higher magnetic field of a  node in the magnetic-field-dependence of the  MQOs \cite{shoenberg_magnetic_1984, bondarenko_first_2001} at 0.58 K, for a field applied along [111]. At this orientation, we identify two prominent oscillation frequencies 341 T and 1399 T (as shown in Fig.~\ref{fig:SmB6_angle_dependence}).  First, in Fig.~\ref{fig:LKfits}a, we illustrate the dependence of the location of the node in oscillatory specific heat on effective mass for a series of effective mass ratios. The location of the node is independent of oscillation frequency, so for clarity, these oscillations are  shown  using a single oscillation frequency of 341 T.  Second, in Fig.~\ref{fig:LKfits}(b), we compare the original data (in black) with the prediction of the L-K model  (in red) for an effective mass ratio of 4.7, this time including oscillations at both 341 T and 1399 T. The overall agreement is  good but limited by the small size of the oscillations in the raw data. Neverthless, inclusion of even just these two most prominent frequencies of oscillation captures much of the observed variation in amplitude with field, confirming our effective mass determination. The size of the decrease in the measured effective mass from 12 T to 24 T at this temperature parallels that seen for the field dependence of $\gamma(H)$.

One cause for caution when measuring flux grown samples is that they often possess Al inclusions, and  torque measurements have shown such inclusions can produce MQOs at frequencies similar to those expected for SmB\textsubscript{6} \cite{thomas_quantum_2019}.   For our flux grown samples, the absence of a discernible jump in the zero-field electronic specific heat of 1\% or greater at the Al superconducting transition temperature of 1.163 K places an upper limit on the actual Al percentage of less than 5\%.  This is a critical test since, if Al inclusions are producing MQOs, then they must arise from high quality crystalline material.  In Fig.~\ref{fig:LKfits}(c) we therefore compare our measurements for the same trace as Fig.~\ref{fig:LKfits}(b) with the corresponding oscillation amplitude and magnetic field dependence expected for an aluminum sample, using the known oscillation frequencies and effective masses of Al \cite{larson_low-field_1967}.  We see here  that even at the 100\% level (pure Al), we are unable to account for the amplitude of the MQOs we see in the specific heat. Further, the effective mass determined from location of the observed node in the amplitude of the  oscillatory specific heat is incompatible with that which would be observed for aluminum, given an effective mass of 0.13 $m_e$. The MQO observed in our samples cannot arise from aluminum inclusions.

 The large effective masses determined above are consistent with a recent first-principles, parameter-free all-electron electronic-structure model for SmB\textsubscript{6}  (\(\frac{m^{\ast}}{m}\) = 2.0 - 22.0 depending on the band) \cite{zhang_understanding_2020}, but are in contrast with values ranging from \(\frac{m^{\ast}}{m}\) = 0.1 - 1.0   found from torque magnetometry  \cite{li_two-dimensional_2014, tan_unconventional_2015, xiang_bulk_2017, hartstein_fermi_2018}. One possible theoretical explanation for the discrepancy in effective mass values observed by specific heat and magnetic torque would be the simultaneous existence of light and heavy quasiparticle masses, as has been proposed for SmB$_6$   \cite{harrison_highly_2018a}.  In this theoretical model, the MQOs arise when a highly asymmetric nodal semimetal forms at low temperature with carriers populated from disorder-induced in-gap states in small-gap Kondo insulators \cite{shen_quantum_2018}. Whether this theory allows the formation of charge neutral excitations is not clear to us but in any case, it would be interesting if the theory were to be extended to include a  calculation of the oscillatory specific heat,  so as to enable a more direct comparison with our results.  Additionally, recent experimental studies on the Kondo insulator YbB\textsubscript{12} suggest a two-fluid picture for the origin of the observed MQO profile in which neutral quasiparticles coexist with charged fermions \cite{Xiang_YbB12}.  In future measurements, we hope to use still higher resolution calorimeters to measure $C(H,T)$ vs $\phi$  to probe for light and heavy effective masses in high quality flux and float-grown samples.

In conclusion we have resolved MQOs in the high field residual specific heat of SmB\textsubscript{6} that show good agreement with theoretical expectations for the dependence of oscillation frequency on crystallographic orientation for SmB\textsubscript{6}, consistent with the existence of neutral quasiparticles at the Fermi surface. 
%even though the parameters needed to describe the observed specific heat oscillations within L-K theory  indicate much larger masses than those previously determined from dHvA measurements. 
It has  been shown in recent theoretical work that neutral quasiparticles arise naturally in mixed valence systems as Majorana excitations \cite{varma_majorana_2020}.  Such excitations would exhibit no charge transport in linear response, but would indeed show MQOs in magnetization, as well as specific heat, consistent with our observations.  Further studies to accurately determine spin-splitting attenuation factors are required to support or oppose these claims.
% In future measurements, we hope to use still higher sensitivity calorimeters to measure $C(H,T)$ vs $\phi$ systematically in high fields to probe for light and heavy effective masses in high quality float- and flux-grown samples at low temperatures, with a goal of the direct observation of oscillations in the specific heat at high magnetic fields.

\begin{ack}

We would like to acknowledge the contributions of Dale Renfrow of the Smith College Center for Design and Fabrication and 
%, who helped design and machine key parts of the calorimeters
Ju-Hyun Park, William Coniglio, and Ali Bangura  of NHMFL for technical support.
% with cryostat and magnet operations at the NHMFL. 
The samples were grown at Los Alamos National Laboratory by P. F. Rosa.  
% This work was supported in part by the Smith College Bull-Paganelli  fund
%endowment for equipment in support of research with undergraduates. 

This work was also supported in part by the U.S. Department of Energy Office of Basic Energy Science, Division of Condensed Matter Physics grant DE-SC0017862 (P.G.L. and A.P.R.).  A portion of this work was performed at the National High Magnetic Field Laboratory, which is supported by the National Science Foundation Cooperative Agreement No. DMR-1644779 and the State of Florida.
\end{ack}

\section*{References}
\bibliography{SmB6main}% Produces the bibliography via BibTeX.

\end{document}